\documentclass[useAMS,usenatbib,a4paper]{mn2e}

\usepackage{aas_macros}

\usepackage{graphicx}
\usepackage{times}
\usepackage{amsmath}
\usepackage{pifont}
\usepackage{mathptm}

\newcommand{\<}{\begin{eqnarray}}
\renewcommand{\>}{\end{eqnarray}} 
\renewcommand{\bar}{\overline}
\renewcommand{\tilde}{\widetilde}
\renewcommand{\hat}{\widehat}
\newcommand{\Msun}{\ensuremath{\rmn{M}_\odot}}

\title[Star formation history of the Large Magellanic Cloud]{
The star formation history of the Large Magellanic Cloud as seen by star clusters and stars}
\author[Th. Maschberger and P. Kroupa]
{Thomas Maschberger$^{1,2}$\thanks{e-mail: tmasch@ast.cam.ac.uk} and Pavel Kroupa$^2$ \\
$^1$ Institute of Astronomy, Madingley Road, Cambridge CB3 0HA\\
$^2$ Argelander-Institut f{\"u}r Astronomie, Auf dem H{\"u}gel 71, D-53121 Bonn, Germany}

\date{13-10-2010}

\pagerange{\pageref{firstpage}--\pageref{lastpage}} \pubyear{2010}

\begin{document}


\maketitle

\label{firstpage}

\begin{abstract}
The aim of this work is to test to what extent the star cluster population of a galaxy can be utilised to constrain or estimate the star formation history, with the Large Magellanic Cloud as our testbed.
We follow two methods to extract information about the star formation rate from star clusters, either using only the most massive clusters \citep[following ][]{maschberger+kroupa2007} or using the whole cluster population, albeit this is only possible for a shorter age span.
We compare these results with the star formation history derived from colour-magnitude diagrams and find good overall agreement for the most recent $\approx 1$ Gyr.
For later ages, and especially during the ``cluster age gap'', there is a deficiency of star clusters in relation to the star formation rate derived from the colour-magnitude diagram.
The star formation rates following from the whole cluster population lie a factor of $\approx$ 10 lower than the star formation rates deduced from the most massive clusters or from the colour-magnitude diagram, suggesting that only $\approx 10$\% of all stars form in long-lived bound star clusters.
\end{abstract}


\section{Introduction}

The understanding of galaxy evolution is a major goal of astrophysics.
Every large-scale event in the life of a galaxy, as e.g. an interaction with another galaxy, has its own pattern of star formation.
Since stars can have long lifetimes, the stellar population preserves information of such events, allowing one to re-trace the galaxy's evolution from the present stellar content.
In this work we focus on the star formation history, the progression of the star formation rate in time.
This study has two main aspects, the comparison of two different methods to obtain a star formation history, from colour-magnitude diagrams and from the star cluster population using the method of \citet{maschberger+kroupa2007}.
Furthermore we discuss the star formation history of the Large Magellanic Cloud, which serves as a ``guinea pig'' for the comparison.

The common method to obtain a detailed star formation history is to observe the stars in a galaxy (or a part of it).
From the distribution of the stars in a colour-magnitude diagram (CMD) the star formation rate at a given time can be derived using modelled tracks of stellar evolution.
To get a result which is representative for the whole galaxy it is necessary to observe a significant fraction of the stars in the galaxy, distributed over a large area. 
This leads to limitations of this method: since individual stars need to be resolved, only nearby galaxies can be examined.
Also, a large number of stars and a large area demand a big observational effort.
Fortunately, the Large Magellanic Cloud has been extensively observed, so that a set 24 million stars is available, from which \citet{harris+zaritsky2009} derive the star formation history.

Another approach to infer a star formation history was presented by \citet{maschberger+kroupa2007}.
Here the fact is used that practically all stars form in star clusters. 
The notion of a star cluster is here taken in a wider sense, denoting stellar assemblies from the smallest size, say a dozen stars, up to classical globular clusters, and does not necessarily imply a bound system.
Whereas low-mass clusters will disperse their stars rather quickly into the galactic field, massive clusters have lifetimes comparable to a Hubble time.
Since the time distribution of massive clusters is related to the star formation rate at their birth, they can be used to find the star formation history of their host galaxy.
This approach using the most massive star clusters has been investigated by \citet{maschberger+kroupa2007} from a theoretical point of view, applying  Monte-Carlo models to study how reliable the massive clusters trace the star formation history.
The value of this method is that, as (massive) star clusters are observable up to much further distances than individual stars, the star formation history of a wider range and number of galaxies can potentially be obtained.

Additionally the time-sequence star formation occurring in star clusters can be determined by simply taking all star clusters into account, given that also a large fraction of lower-mass star clusters is usually observed.
\citet{subramaniam2004} compared in six regions of the Large Magellanic Cloud the cluster formation rates and star formation rates (derived from colour-magnitude diagrams), and found that in general the number of formed clusters follows the ``field'' star formation rates.
Cluster formation rates only have also been derived by \citet{girardi-etal1995}, \citet{pietrzynski+udalski2000}, \citet{hunter-etal2003}, and \citet{degrijs+anders2006}
Our approach is slightly different, we use the total mass in clusters per time instead of number of clusters per time.
With the diminishing brightness of star clusters as they age the observed number of older star clusters decreases, so that only a shorter fraction of the galaxy's life-time can be investigated in this way.

The Large Magellanic Cloud has been the target of many research projects and is ideal to compare the CMD and star cluster methods.
In the literature there are a number of studies available on the star formation history of the Large Magellanic Cloud, which are based on the CMD approach \citep[e.g.][]{harris+zaritsky2009, olsen1999,holtzman-etal1999,dolphin2000,smeckerhane-etal2002,subramaniam2004,javiel-etal2005}.
Furthermore, the star cluster population of the Large Magellanic Cloud has been investigated, and ages and masses of a large fraction of the star clusters have been determined \citep{pietrzynski+udalski2000,hunter-etal2003,degrijs+anders2006}.
This enables us to study the star formation history of a galaxy with two independent methods.

These introductory remarks outline the structure of this work, in summary:
After first discussing the star cluster data set, we derive the star formation history of the Large Magellanic Cloud from the most-massive clusters (Sec \ref{sec_mostmassive}) and from the total population (Sec. \ref{sec_total}).
Then we describe the results obtained from colour magnitude diagrams (Sec. \ref{sec_cmd}).
We finish with a comparison of the results (Sec. \ref{sec_comparison}) and a summary (Sec. \ref{sec_summary}).

\begin{figure}
\includegraphics[width=8cm]{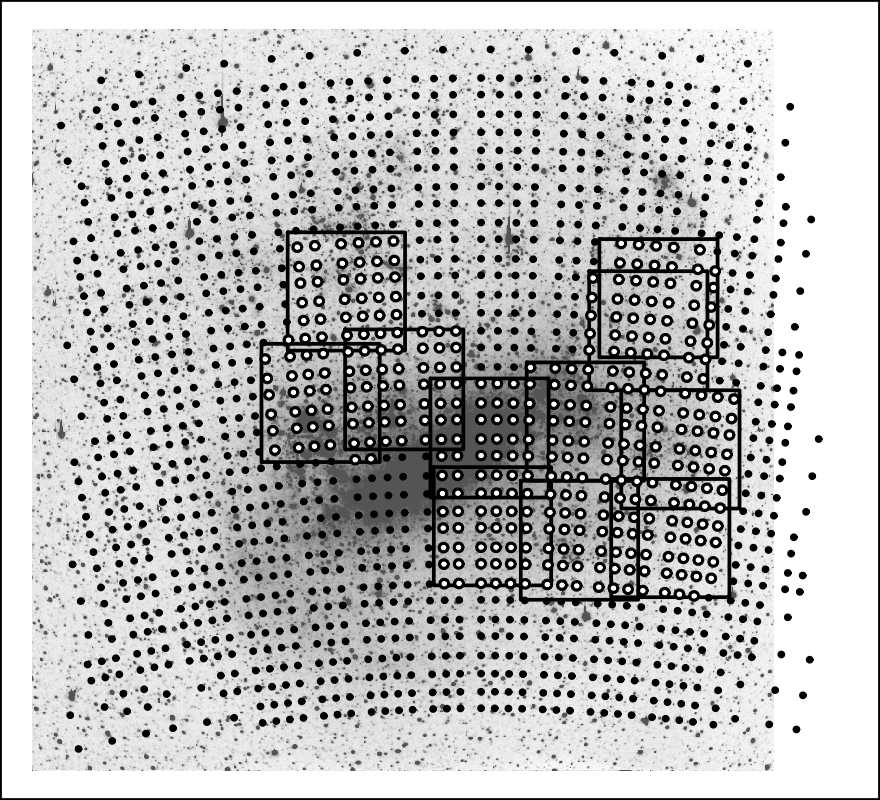}
\caption{\label{selection}
Overview of the observed regions in the Large Magellanic Cloud (background image from \citet{bothun+thompson1988} with astrometry by \citealp{parker-etal1998}).
The dots mark the centres of the fields observed by \citet{harris+zaritsky2009} and for which a star formation history was derived using a colour-magnitude diagram.
The squares are the boundaries of the regions observed by \citet{massey2002}, in which \citet{degrijs+anders2006} derived star cluster ages and masses.
The open dots are the fields of \citet{harris+zaritsky2009} which we selected for comparison.
}
\end{figure}

\section{The star formation history of the Large Magellanic Cloud as seen by star clusters}

\begin{figure}
\includegraphics[width=8cm]{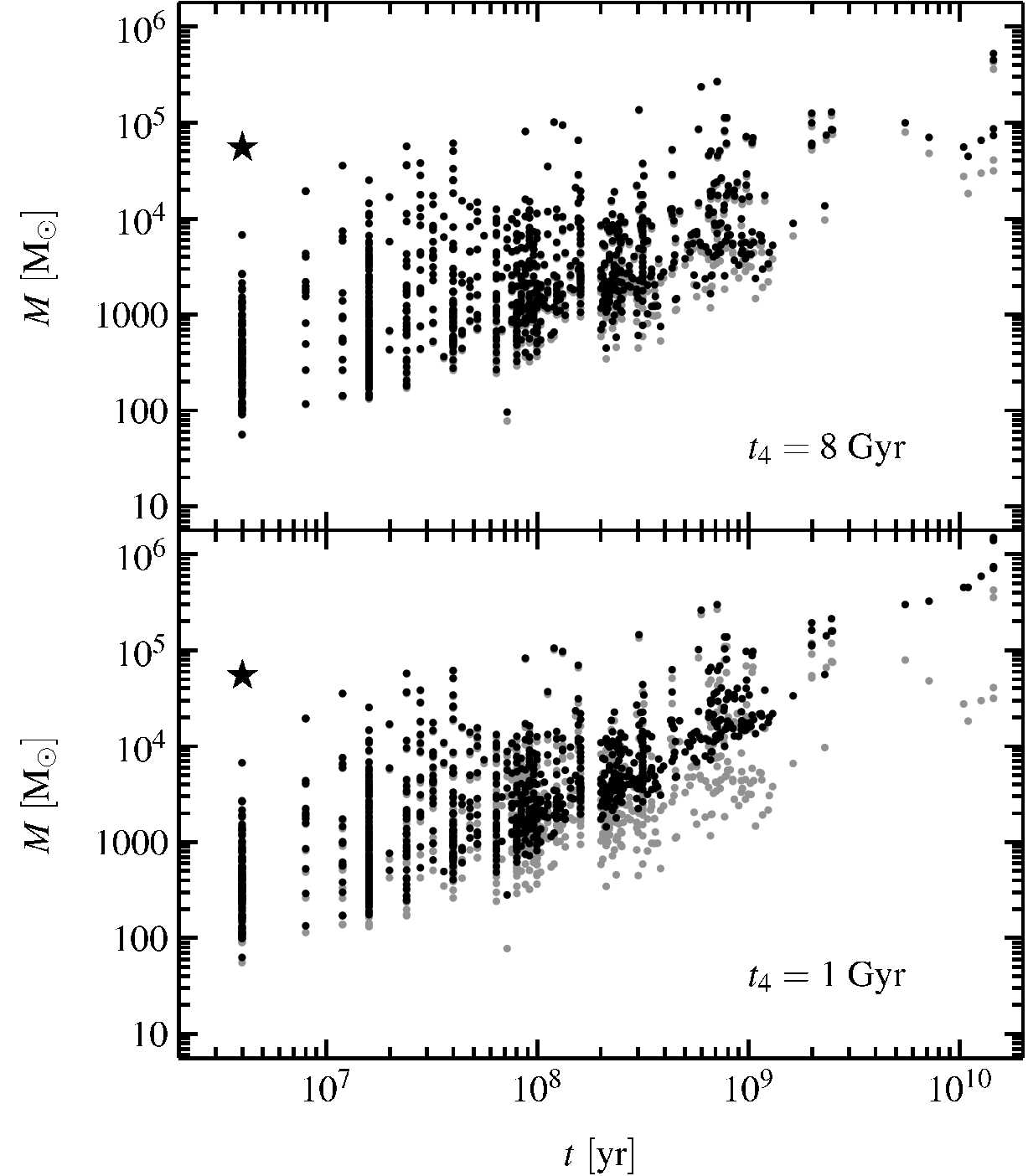}
\caption{\label{lmc-clusters}
Age-Mass diagram of the star clusters in the Large Magellanic Cloud\citep[grey circles original data from][]{degrijs+anders2006}.
The cluster masses have been corrected for dynamical evolution using eq. \ref{lamersevolve} with $t_4=$ 8 Gyr (black dots, top panel) and $t_4=$ 1 Gyr (black dots bottom panel).
30Dor, not part of the original data set, is shown as a star.
}
\end{figure}

\subsection{Data}

The ages and masses of the star clusters we use for the analysis are taken from \citet{degrijs+anders2006}, which re-analysed the photometry of \citet{hunter-etal2003}, which itself is based on the observations by \citet{massey2002}.
The rectangles in Fig. \ref{selection} show the spatial coverage of the observed regions with star clusters.
The star cluster ages and masses were derived by \citet{degrijs+anders2006} from broad-band spectral energy distributions using their AnalySED tool which is based on the {\sc galev} single stellar population models \citep{kurth-etal1999,anders+fritze2003,anders-etal2004}.
The age uncertainties for the 922 star clusters are in the range $\Delta \log_{10} (\tau / \mathrm{yr}) \le 0.35$ \citep{degrijs-etal2005}.
Due to discrete isochrones the age-mass diagram in Fig. \ref{lmc-clusters} shows columns of star clusters of the same age.
The lower mass limit for detection of clusters increases with increasing cluster age, as clusters fade due to stellar evolution and dynamical loss of stars, leading to the wedge-like shape of the data in Fig. \ref{lmc-clusters}.

The AnalySED tool provides ``initial'' masses of the star clusters which are corrected for mass loss due to stellar evolution.
However, the mass of a star cluster diminishes in time also because stars are lost in consequence of dynamical evolution.
As we need true initial masses for the star clusters we correct for the dynamical evolution using the formulae of \citet{lamers-etal2005a}.
Given the dissolution time of a $10^4$\ \Msun\ star cluster, $t_4$, the initial star cluster mass of age $t$ is given as
\< M_\mathrm{ini} &=& \left[  \left( \frac{M(t)}{\Msun} \right)^{0.62} + \frac{0.62 \times t}{ (t_4/660)^{1.034} }\right]^{1.61} \label{lamersevolve} \>
(This follows from  combining eqq. 7 and 11 of \citet{lamers-etal2005a}, with $\gamma=0.62$ and omitting the term for stellar evolution in eq. 11).
For the Large Magellanic Cloud \citet{boutloukos+lamers2003} found $\log_{10} t_4=9.7$ (using a smaller data set) and \citet{degrijs+anders2006} gave the slightly larger value of $\log_{10} t_4 = 9.9$ (8 Gyr).
\citet{parmentier+degrijs2008} carefully performed a reanalysis of the dissolution time and concluded that it is with the current data set only possible to constrain $t_4$ to be larger than 1 Gyr.
Therefore we use two values for $t_4$, 1 Gyr and 8 Gyr, to correct for dynamical evolution.
To visualise the difference between these values we show in Fig. \ref{lmc-clusters} the not back-evolved masses as grey circles, and the ``true'' initial cluster masses as dots, using $t_4 = 8$ Gyr in the top panel and $t_4 = 1$ Gyr in the bottom panel.
For us a larger $t_4$ seems to be more realistic, as the Small Magellanic Cloud has a similar value \citep[$\log_{10} t_4 = 9.9$, ][]{lamers-etal2005}, and more massive spiral galaxies with a deeper gravitational potential have smaller values.
We will, however, discuss below the implications of both values when determining the star formation history from the most massive clusters.

Further features in the age-mass diagram besides the typical wedge-like shape were pointed out by \citet{degrijs+anders2006}:\\
(1) The large densities of clusters at $\log_{10} \tau $ of 6.6 and 7.2: These are caused by the fitting procedure. 
There are no isochrones for clusters younger than 4 Myr ($\log_{10} \tau = 6.6$), and at $\log_{10} \tau = 7.2$ the isochrones are discrete due to rapid evolution. 
This does not have a large influence on the determined SFH.\\
(2) The under-density of data points between $\approx$ 3 Gyr and 13 Gyr ($\approx 9.5 \le \log_{10} \tau \le 10.1$), which is the ``well-known LMC cluster age-gap''.\\
(3) Overdensities at $7.8 \le  \log_{10} \tau \le 8.0$, $2.8 \le \log_{10} (M/\Msun) \le 3.4$  and $ 8.2 \le \lg \tau \le 8.4 $, all masses. 
These features could be caused by the last encounters between the Large and Small Magellanic Cloud, but this cannot be concluded with sufficient certainty because of the lack of better age resolution and lack of orbital information for the galaxies. 

It has also to be noted that this star cluster sample does not contain the 30Dor region, containing the young star cluster R136.
It was classified as a newly formed star cluster (``NC'') by \citet{bica-etal1999} and so in a group of objects which were not selected by \citet{hunter-etal2003}.
However, R135 is a massive star cluster having a mass of $\approx 5.5 \times 10^4\ \Msun$ \citep{hunter-etal1995}, and is the most massive star cluster recently formed.
The inclusion of this cluster is therefore crucial to the method used in the next Section.

\subsection{Star Formation History using the most massive star clusters}\label{sec_mostmassive}

In \citet{maschberger+kroupa2007} we presented and tested a method to derive the star formation history of a galaxy using the most massive clusters.
This method is based on the observation that the brightness of the brightest young cluster in a galaxy is correlated with the (present) star formation rate \citep{larsen2002,weidner-etal2004,bastian2008}.
This can be understood following the argument of \citet{weidner-etal2004}.
Within a certain time span of the galaxy's lifetime, $\delta t$, the amount of mass assembled in (long-lived) stellar clusters is proportional to the star formation rate,
\< M_\mathrm{clusters} = A \ \mathrm{SFR}\  \delta t \label{sfrdeltat} \>
($A$ is the proportionality constant).
This mass in clusters is related to a number of clusters that have formed,
\< N_\mathrm{clusters} = \frac{M_\mathrm{clusters}}{\bar{M}},\>
where a universal cluster mass function is assumed to calculate the average mass of a star cluster, $\bar{M}$. 
Interpreting the star cluster mass function as a probability distribution, this then allows one to calculate the distribution of the most massive star cluster, $M_\mathrm{max}$, that would be expected for the given $N_\mathrm{clusters}$ .
From this model follows a relation of the mass of the most massive star cluster with the star formation rate within $\delta t$.
This can be inverted to $SFR=f(M_\mathrm{max})$ which can be used to determine the star formation rate over time, discretised by $\delta t$.

In general $M_\mathrm{max}$ follows a probability distribution, related to the star cluster mass function, which has to be taken into account for the inversion (details of this can be found in \citealp{maschberger+kroupa2007}).
To minimise the number of assumptions, especially the exact form (pure power law or Schechter function as suggested by \citealp{gieles-etal2006b}) and parameters of the cluster mass function, we use the relation of the mean mass of the most massive cluster and the star formation rate.
The $\bar{M}_\mathrm{max}$--SFR relation can be directly calibrated with the observed relation of the brightest young cluster and  the star formation rate in a galaxy (assuming that the brightest cluster is also the most massive one of the most recent time interval, an assumption which is discussed in more detail below).
This $\bar{M}_\mathrm{max}$--SFR relation  is then applied to a mean mass of the observed most massive clusters over several $\delta t$ (choosing the number of used $\delta t$ such that during the whole time of averaging the star formation rate in the galaxy is not changing significantly).
By using a moving averaging window (moved in steps of $\delta t$) the time resolution of the obtained star formation history can be increased.
The length of the averaging window is essentially constrained by the age uncertainties of the star clusters, which are constant in logarithmic space, so that we keep the averaging window also constant in $\log_{10}$.

By using $\bar{M}_\mathrm{max}$ and the empirical calibration we have avoided the need of the exact knowledge of the star cluster mass function.
However, another crucial ingredient in this method is the formation epoch, $\delta t$, which needs more explanation.
In this context the often mentioned ``size-of-sample'' effect has to be discussed.
The ``size-of-sample'' effect is simply the statistical increase of the mass of the most massive cluster with increasing sample size.
With the general assumptions of an unchanging cluster mass function and constant cluster formation rate (number per time) a logarithmic age-mass diagram has the characteristic upper envelope of an increasing mass with time.
Equally-spaced time intervals in logarithmic space contain more physical time, thus more clusters are formed and subsequently the mass increases.
However, this is not the full picture as the cluster (or star-) formation rate can change with time, leading for example to the ``age gap'' in the Large Magellanic Cloud where barely clusters are found.
A mathematically more correct description would be a star cluster mass function depending on both mass {\it and} time, which is however not very practical.
Here the formation epoch, $\delta t$, comes into the play: this is the time by which the time evolution of a galaxy is discretised.
With a reasonable choice of $\delta t$ the star formation rate in the galaxy can be assumed to stay constant, simplifying the statistical treatment.
The increasing envelope in the log(age)-log(mass) diagram is preserved with using $\delta t$ \citep[shown in fig. 3, top panel, of ][]{maschberger+kroupa2007}.
The difference to the established understanding of the ``size-of-sample'' effect is that one does not increase the size of a single sample, but one instead increases the number of samples.

The question is now what a reasonable size for $\delta t$ is. 
Already mentioned was the need for the star formation rate to be constant over $\delta t$, which gives an upper limit for $\delta t$ of $\approx$ 100 Myr, the dynamical time of a galaxy.
In fact, the star formation rate should be constant over several $\delta t$ so that it can be averaged over several $M_\mathrm{max}$.
We follow here the choice of \citet{weidner-etal2004} and \citet{maschberger+kroupa2007} of 10 Myr, for the practical reason that most clusters are of this age in the observational $M_\mathrm{max}-\mathrm{SFR}$ plot, as the luminosity of a star cluster peaks at about 10 Myr.
Thus the brightest cluster is in many cases of this age and at the same time the most massive.
With a different choice of $\delta t$ the brightest cluster would have to be replaced by the actual most massive cluster in the normalisation, i.e. a true $M_\mathrm{max}$-SFR diagram would have to be observed.
One possible interpretation of $\delta t = 10$ Myr would be that it is the typical time-scale on which the inter-stellar medium rearranges itself into a coeval population of star clusters that are distributed according to the star-cluster initial mass function \citep[cf. ][]{weidner-etal2004}.
Anyway, the comparison of the star cluster results with the CMD results will give an indirect check if our normalisation is correct.

\begin{figure}
\includegraphics[width=8cm]{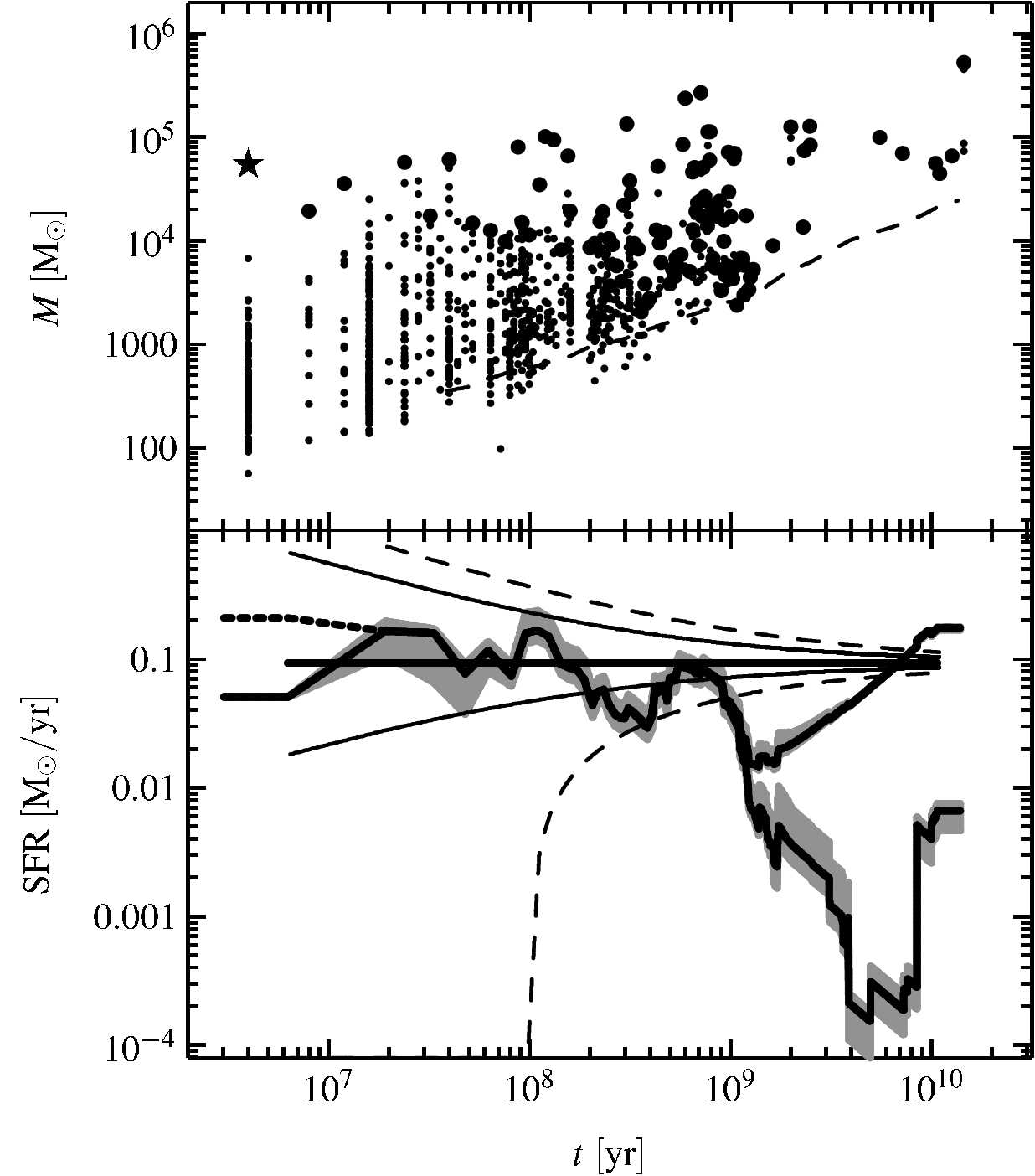}
\caption{\label{sfhmmax}
Star formation history of the Large Magellanic Cloud derived from the most massive clusters.
In the top panel we shows the age-mass diagram of the star clusters (with dynamically back-evolved masses, $t_4$=8 Gyr), including 30Dor as the star and highlighting the most massive clusters of each formation epoch with bigger circles.
The dashed line is the fading limit.
The lower panel shows the star formation history derived from the most massive clusters (with back-evolved masses and $t_4$=8Gyr) as thick solid lines, splitting into two branches, the upper using the fading limit mass in gaps and the lower using $M_\mathrm{max}=0\Msun$.
The thick dashed branch at early ages follows with including 30Dor.
The gray area is the uncertainty propagated from the uncertainty in the cluster masses.
A horizontal line marks the average star formation rate and is enclosed by the $1\sigma$ and $2\sigma$ curves (solid and dashed) derived from the statistical spread of the $M_\mathrm{max}$.
}
\end{figure}

The method of going in 10-Myr-steps through age to estimate the star formation rate only works if there are star clusters in each formation epoch.
However, an inspection of the age-mass diagram (Fig. \ref{lmc-clusters})  shows that for ages larger than 1 Gyr there are long intervals without clusters.
For a formation epoch in which no cluster is detected a direct estimate of the star formation rate is not possible.
But a non-detection does one allow to estimate limits for the star formation rate.
If no cluster is detected, then none has formed massive enough to be detected, which implies a low star formation rate.
By using a star cluster evolution model the observational limiting magnitude can be translated into a mass, the fading limit.
As no star cluster is detected above the fading limit, all cluster that might have been formed in this formation epoch must have had smaller masses.
Thus the fading limit can be used as an upper limit for the most massive cluster in this epoch, and with this mass an upper limit for the star formation rate can be calculated.
The lower limit for the star formation rate in an empty epoch is no star formation.

This treatment of empty formation epoch gives us two results for the star formation history, an upper and a lower limit.
When the upper and lower limit are identical (i.e. when only ``full'' formation epochs are used), the most massive cluster method (using the $\bar{M}_\mathrm{max}$-SFR relation) gives an estimate of the star formation history, otherwise the star formation rate can only be constrained by an upper and lower limit.
This is an implicit quality assessment of the method.
Further uncertainties of results are caused by the probabilistic nature of $M_\mathrm{max}$, which mainly affects younger ages, and the uncertainties in the masses.
We discuss these and their treatment after presenting the results for the Large Magellanic Cloud.

Besides the availability of data there are two other sources of uncertainty for the most-massive cluster method, statistical scatter and the age/mass uncertainties.
Due to the probabilistic nature of $M_\mathrm{max}$ the stochastic scatter in the recovered star formation rate is very large for young ages, as averaging occurs only over a few $\delta t$, and decreases with increasing time.
The amount of stochastic scatter has been determined from Monte-Carlo experiments by \citet[][their sec. 4.3, eqq. (16) and (17)]{maschberger+kroupa2007}.
The age uncertainties are accommodated for by the averaging window, which is kept constant in logarithmic time and has approximately the size of the age uncertainties (0.5 dex).
The mass uncertainty is propagated to an uncertainty in the star formation rate by using the $M_\mathrm{max}$ values plus/minus their uncertainty in mass.

In Fig. \ref{sfhmmax} we show in the upper panel the age-mass diagram of the star clusters, where the clusters identified as $M_\mathrm{max}$ are the large dots.
R136 is shown as an open circle, as it is not contained in the \citet{degrijs+anders2006} sample.
The dashed line is the fading limit, the mass that a cluster with the lowest observed brightness would have (calculated with the {\sc galev} models).
The upper and lower limit for the star formation history are shown as bold lines, which have the same values up to $\approx$ 1 Gyr (calculated with $t_4 = 8$ Gyr).
For the youngest ages the dashed branch follows by including R136 in the star cluster sample.
The uncertainty in the star formation rate introduced by the uncertainties in the cluster masses is visualised as a grey region.
To assess the significance in variations of the star formation rate we show a constant star formation rate (the thick solid line at $\approx 0.1 \Msun/\mathrm{yr}$) which is embraced by the statistical $1 \sigma$ and $2 \sigma$ scatter (thin solid lines and dashed lines, see \citealp{maschberger+kroupa2007}).

Generally the obtained star formation history follows the distribution of the star clusters for about one Gyr, when the number of clusters starts thinning out.
Both the upper and lower limit agree for that period, so that the result of the method should be an estimate of the star formation history until that age.
The peaks in the star formation rate are somewhat displaced when compared to the loci of the massive clusters which is caused by the time averaging.
For ages younger than 100 Myr it is not possible to establish whether the variations in the derived star formation rate are caused by variations in the actual star formation rate of the Large Magellanic Cloud because of the large stochastical scatter.
In the age range from 100 Myr to 1 Gyr the derived star formation rate deviates $\approx 2 \sigma$ from a constant star formation rate, which should be caused by a decrease in the actual star formation rate.

\begin{figure}
\includegraphics[width=8cm]{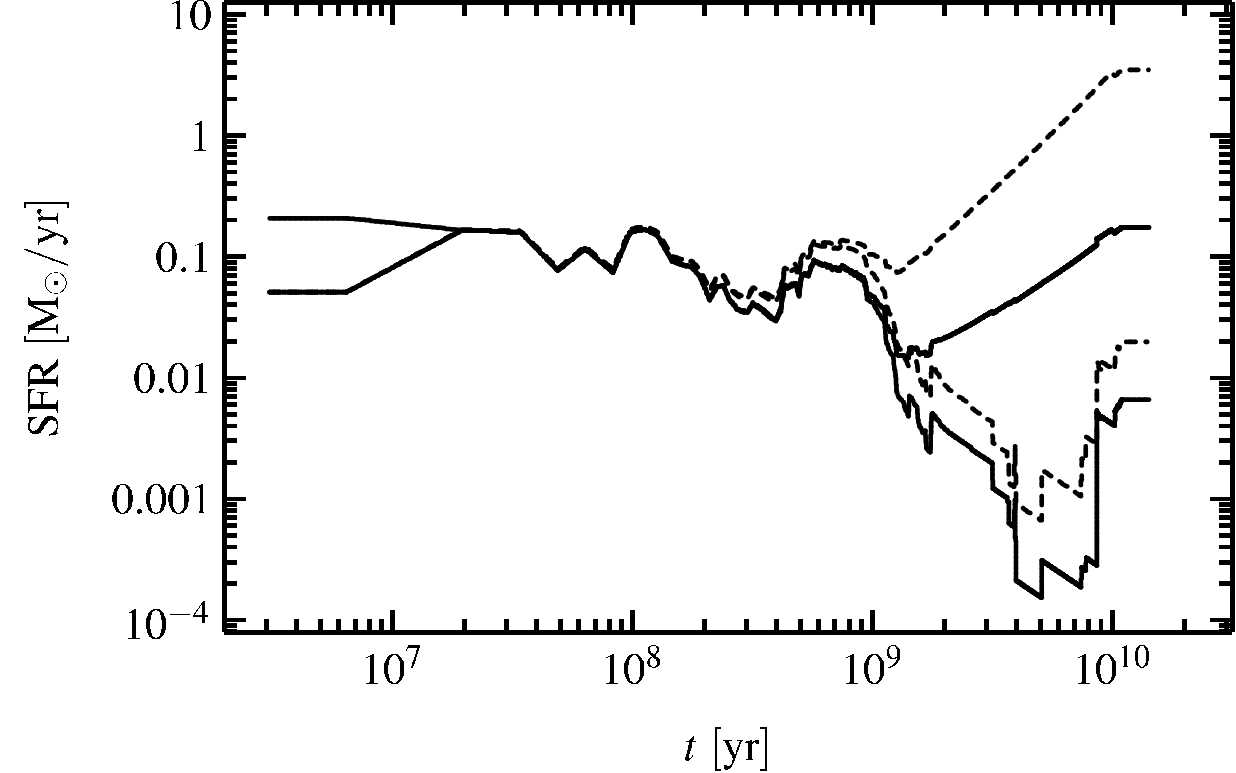}
\caption{\label{sfhmmax_t4}
Influence of $t_4$ on the star formation history of the Large Magellanic Cloud using star clusters.
For the solid line $t_4=8$ Gyr and for the dashed line $t_4=1$ Gyr was used. 
}
\end{figure}

We now turn to discuss the effects of the two values for  $t_4$.
Figure \ref{sfhmmax_t4} shows this, with the star formation history using $t_4 = 8$ Gyr as the solid line and $t_4 = 1$ Gyr as the dashed line.
Our results use only massive clusters, which are significantly affected by dynamical evolution only after a long time.
Therefore the two solutions for the star formation history differ only at large ages.
Essentially, the shorter $t_4$ implies a stronger dynamical evolution of the clusters, which consequently had larger initial masses, leading to a higher derived star formation rate.
The differences in the star formation histories for ages younger than $\approx$ 1 Gyr are only small.
For larger ages the star formation rates are by a factor of $\approx$ 10 larger for the smaller $t_4$.

\subsection{Star Formation History using the total mass in star clusters}\label{sec_total}

\begin{figure}
\includegraphics[width=8cm]{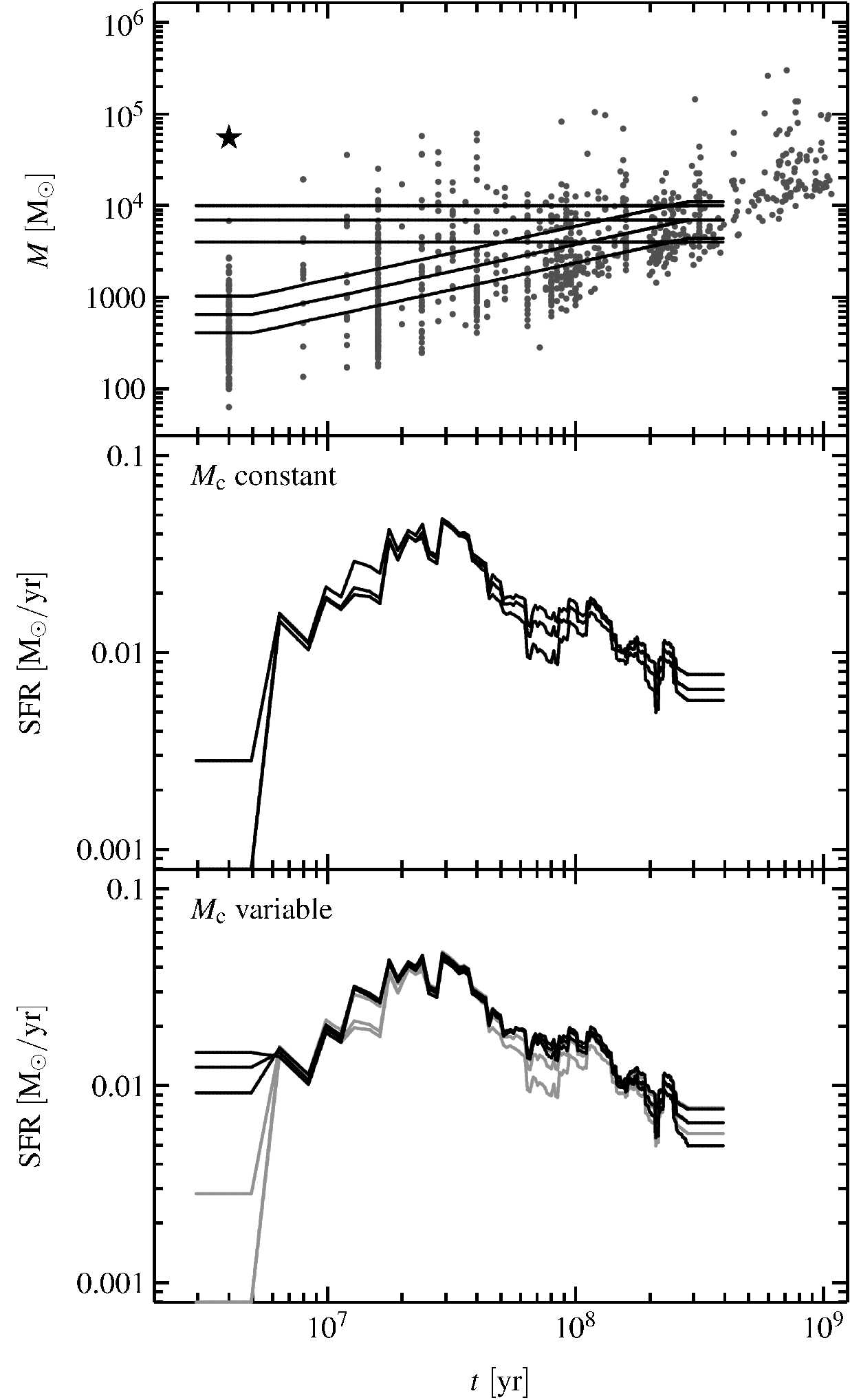}
\caption{\label{sfhdirect}
Recent history of star formation in star clusters, derived by adding up all cluster masses above a completeness mass, $M_c$ and re-normalised as described in the text (cluster masses were dynamically back-evolved using $t_4$=8 Gyr).
R136, shown as a star, is not included in the analysis.
Various values for $M_c$ have been chosen to demonstrate the dependence of the result on the completeness.
The top panel shows the location of the constant (4000, 7000 and 10000 \Msun) and time-variable $M_c$ (407, 646 and 1023 \Msun at $t=0$) in the cluster age-mass diagram.
For the constant $M_c$ the star formation histories are shown in the middle panel.
The bottom panel contains the star formation histories for time variable $M_c$, also showing the results from the middle panel as grey lines for comparison.
}
\end{figure}

As the Large Magellanic Cloud is very near to the Milky Way not only high-mass but also intermediate-mass clusters (with masses down to a few thousand \Msun) can be detected over an extended time span.
This allows us to use not only the most massive clusters to derive star formation rates, but also the whole cluster population.
The amount of star formation in star clusters in a given time interval is simply the ratio of the total mass of star clusters and the length of the interval,
\< \tilde{\mathrm{SFR}} (t, M_\mathrm{c})  &=& \frac{1}{\Delta t} \sum_{
\substack{ M_i > M_\mathrm{c}\\ t (M_i) \in \Delta t } } M_i, \>
where $M_i$ is the mass of the $i-th$ cluster with its age $t(M_i)$ and $M_\mathrm{c}$ is the completeness mass which follows from the detection limit.

We derive the history  of star formation in star clusters by moving a time interval of constant logarithmic size (0.35 dex) in steps of 1 Myr until its boundary reaches an age of 400 Myr.
For older ages the number of star clusters is too small to reach reasonable results.
We use dynamically back-evolved cluster masses, $M_i$, with $t_4$=8Gyr.

Because the observations do not reach down to the lowest masses which star clusters can have, $\tilde{\mathrm{SFR}}$ gives only a fraction of the total star formation rate in star clusters.
Thus the mass in star clusters has to be extrapolated to a total mass of stars in star clusters by assuming a star cluster mass function, $\zeta \propto M^{-\beta}$, a power law parametrised by an exponent ($\beta$) and a lower and upper mass limit ($M_\mathrm{l}$ and $M_\mathrm{u}$).
$\zeta$ is here normalised as a probability density, i.e.  $\int_{M_\mathrm{l}}^{M_\mathrm{u}} \zeta \mathrm{d} M = 1$.
The normalisation factor follows from the observed fraction of star clusters,
\< \frac{M_\mathrm{obs}}{M_\mathrm{tot}} &=& \frac{ \int_{M_\mathrm{c}}^{M_\mathrm{u}} M \zeta \mathrm{d} M}{\int_{M_\mathrm{l}}^{M_\mathrm{u}} M \zeta \mathrm{d} M},\>
as
\< a(M_c) &=& \frac{\int_{M_\mathrm{l}}^{M_\mathrm{u}} M \zeta \mathrm{d}M}{\int_{M_\mathrm{c}}^{M_\mathrm{u}} M \zeta \mathrm{d} M}. \>
The correct star formation rate in clusters is then
\< \mathrm{SFR} (t,M_\mathrm{c} ) &=& a(M_\mathrm{c} ) \tilde{SFR} ( t,M_\mathrm{c}).\>
The derived star formation history depends on the chosen completeness mass and parameters of the star cluster mass function.

In order to explore the robustness of the obtained results we turn first to the completeness mass, as it seems not to be too well constrained in our data set (see e.g. the discussion in \citealp{parmentier+degrijs2008} and \citealp{maschberger+kroupa2009})
To circumvent this problem we choose various possibilities for $M_\mathrm{c}$, shown as lines in the top panel of Fig. \ref{sfhdirect}, the age-mass diagram.
For a minimal dependence on the parameters of the cluster mass function we choose $M_c$ constant in time, with different values.
Therewith one prohibits systematic effects in the shape of the star formation history caused by wrong parameters for the cluster mass function as the normalisation factor is constant for all age bins.
The results are shown in the middle panel of Fig. \ref{sfhdirect}, for a better comparison re-scaled such that the star formation histories lie near together (actually using in all cases $\beta=2$, $M_\mathrm{l}=100\ \Msun$ and $M_\mathrm{u}=10^6\ \Msun$).
The general structure is the same for all star formation histories, except for the very youngest ages because of a lack of massive clusters.
Further, small variations appear around $\approx 15$ Myr, $\approx$ 60--90 Myr and after $\approx$ 300 Myr.
The first two small variations are caused by an insufficiently small number of clusters more massive than $M_c$.
The discrepancy after 300 Myr is almost certainly caused by a too low mass for the lowest $M_c$, leading to an incomplete data set at these ages.

For an optimal use of the available data we choose a time-variable $M_c$, running parallel to the lower envelope of the star clusters in Fig. \ref{sfhdirect}, top panel.
Now each age bin has an individual normalisation constant, potentially introducing time-dependent systematics.
The lower panel of Fig. \ref{sfhdirect} shows the obtained star formation histories with the results for the constant $M_\mathrm{c}$ (i.e. the results of the middle panel) plotted in grey for comparison.
The overall structure of the star formation history is the same as for constant $M_c$, with the exception that the peak at 90--150-Myr is more like a plateau. 
The small variations at $\approx 15$ Myr, $\approx$ 60--90 Myr disappear with the larger number of clusters used, but the feature at $\approx$ 300 Myr is still present.
The agreement for the different choices of $M_c$ is better than for constant $M_c$, which is rather surprising as the influence of the star cluster mass function is changing over time.
This indicates that the results are robust and no systematical effects are introduced by the time-variable $M_\mathrm{c}$.

\begin{figure}
\includegraphics[width=8cm]{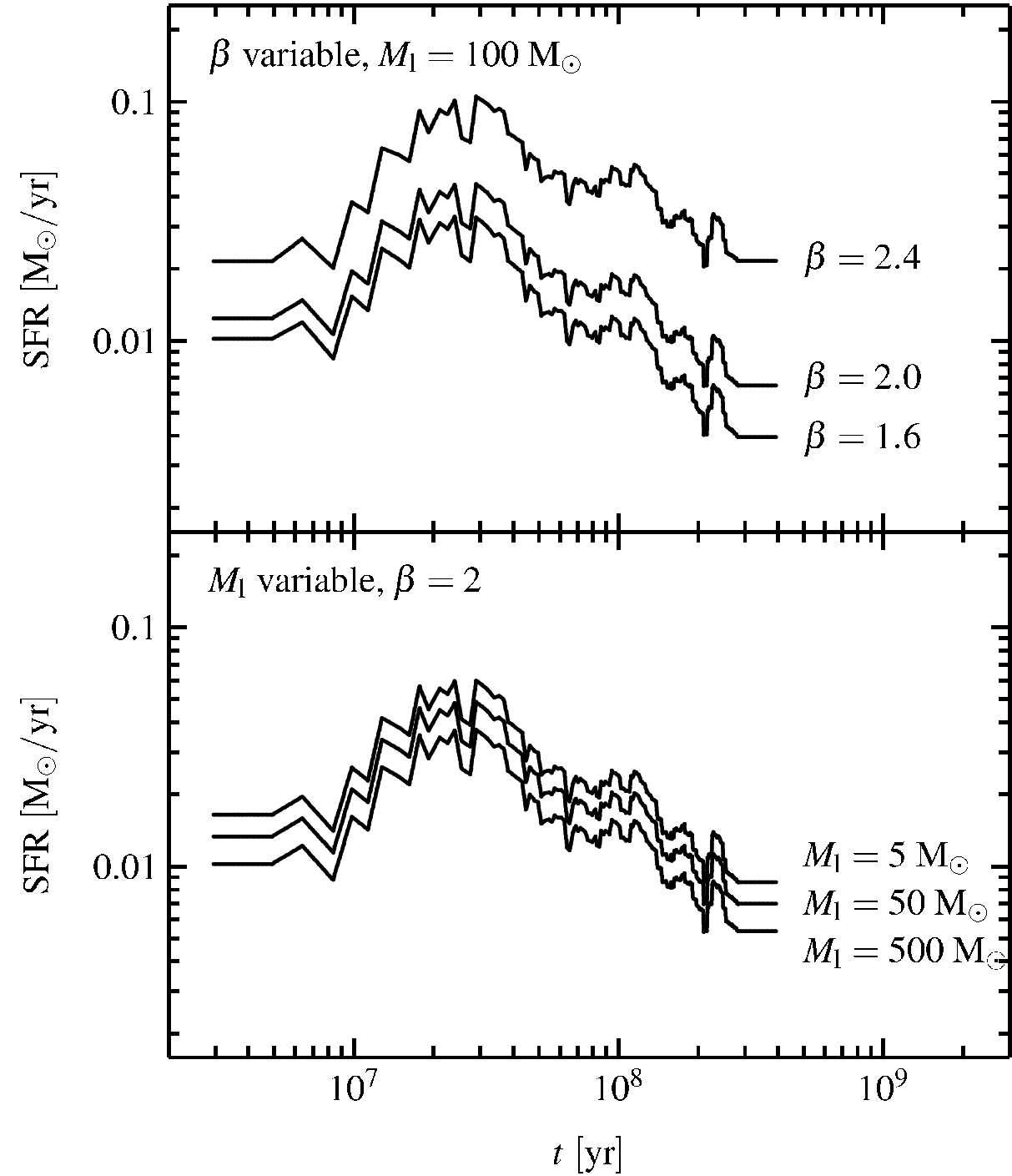}
\caption{\label{sfhdirect_parameters}
Influence of the parameters of the star cluster mass function on the normalisation of the  star formation history, derived from the complete cluster population (dynamically back-evolved cluster masses with $t_4=$ 8 Gyr and time-variable completeness mass $M_c$, the middle in Fig. \ref{sfhdirect}, top panel).
The top panel shows the changes caused by different exponents $\beta$ of the cluster mass function, especially on the older-age history.
The choice of the lower limit $M_l$ (bottom panel) has only minor influence on the absolute value of the star formation rates.x
}
\end{figure}

For the correction of $\tilde{\mathrm{SFR}}$ we used $\beta=2$, $M_\mathrm{l}=100\ \Msun$ and $M_\mathrm{u}=10^6\ \Msun$.
These values, especially $\beta$, are chosen such that the different $M_\mathrm{c}$ all lead to the same result.
As there are various values reported for $\beta$ in the literature (e.g. \citealp{maschberger+kroupa2009}, \citealp{gieles2009} and references therein and \citealp{weidner-etal2004}) we 
show in the top panel of Fig. \ref{sfhdirect_parameters} star formation histories corrected with different values of $\beta$ ($1.6$, $2.0$ and $2.4$).
$M_\mathrm{c}$ is variable in time, starting with 260\ \Msun (which is the second to lowest line in the top panel of Fig. \ref{sfhdirect}).
For larger $\beta$ the fraction of star clusters below $M_\mathrm{c}$ increases, so that the star formation histories start at higher star formation rates.
The increase of the star formation rates for different $\beta$ is also time-dependent for time-variable $M_\mathrm{c}$, so that for older ages the amount by which the star formation rates are corrected increases.
This leads to the growing difference between the curves in Fig. \ref{sfhdirect_parameters}, top panel.
The overall structure, however, remains the same within our range of $\beta$, and no additional features are introduced.

The lower limit of the cluster mass function only has minor influence on the absolute level of star formation histories, as evident in the lower panel of Fig. \ref{sfhdirect_parameters}.
Here we varied $M_\mathrm{l}$, using $M_\mathrm{l}=5\ \Msun$, $50\ \Msun$ and $500\ \Msun$ ($\beta=2.0$ and $M_\mathrm{u}=10^6\ \Msun$).
The star formation rates are a factor of $1.6$ higher for $M_\mathrm{l}=5\ \Msun$ compared to their values for $M_\mathrm{l}=500\ \Msun$.

\section{The star formation history derived from colour-magnitude diagrams}\label{sec_cmd}

\begin{figure}
\includegraphics[width=8cm]{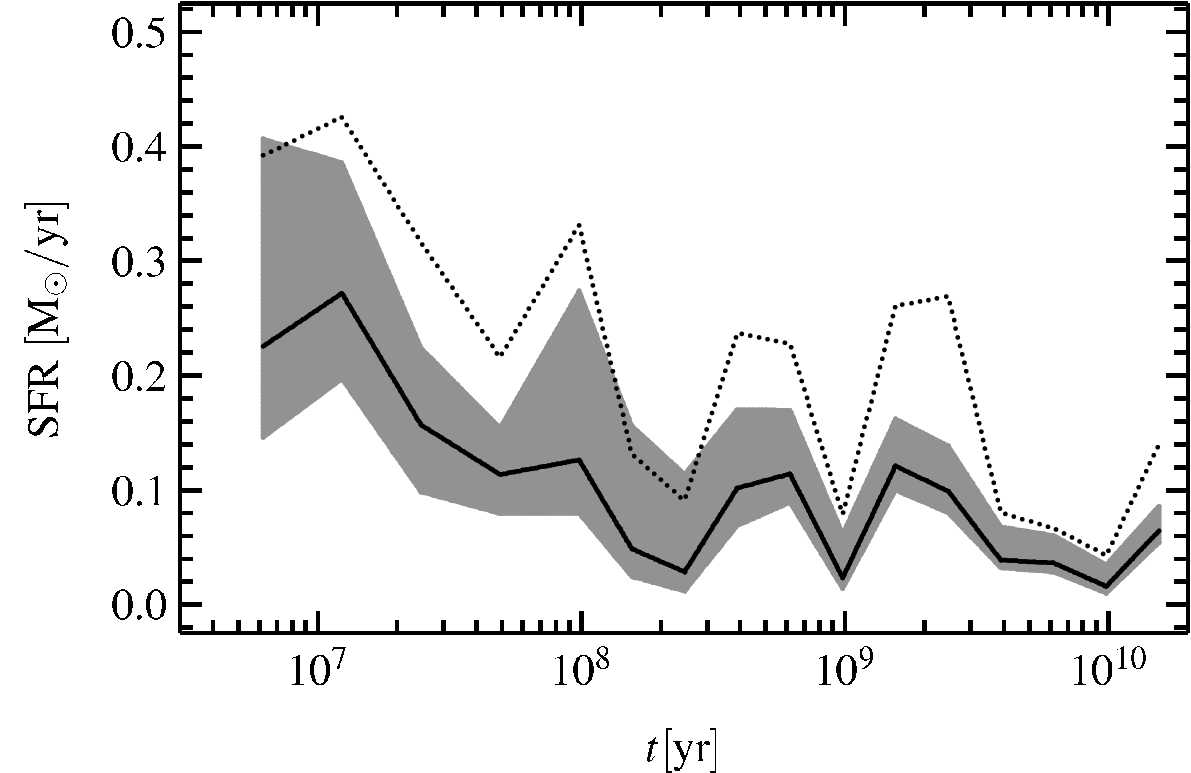}
\caption{\label{sfh-hz}
Star formation history derived by \citet{harris+zaritsky2009} using the colour-magnitude diagram method.
The solid line is the SFH for the regions with in the Massey fields only (open circles in Fig. \ref{selection}), with the uncertainty given by the grey region.
The dotted line above is for the whole of the Large Magellanic Cloud.
}
\end{figure}

For the comparison of our star cluster results with the results utilising colour-magnitude diagrams we use the work of \citet{harris+zaritsky2009}.
They presented the star formation history derived with the StarFISH software  \citep{harris+zaritsky2001} in a grid of fields covering the whole Large Magellanic Cloud (the coverage is shown in Fig. \ref{selection}).
Their photometric catalogue contains 24 million objects, so that each of the individually analysed fields contains some $10^4$ stars.
For the synthetic colour magnitude diagrams the isochrones of the Padova group were used \citep{girardi-etal2002}.
The temporal resolution of the star formation history is given by age bins of 0.3 dex size for ages younger than 100 Myr and bins of 0.2 dex for older ages.
Solutions for the star formation history were obtained for four metallicities, $Z=0.001$, $Z=0.0025$ (interpolated and used only for ages larger than 100 Myr), $Z=0.004$ and $Z=0.008$.

As their photometry does not reach the main sequence turnoff point for the old population the extraction of the early star formation history was difficult.
Therefore \citet{harris+zaritsky2009} restricted StarFISH to fit only a single age bin covering all ages older than 4 Gyr in the bar region.
Within the bar region they used the typical star formation history from solutions for the star formation history derived using HST data (\citealp{olsen1999}, 
\citealp{holtzman-etal1999} and \citealp{smeckerhane-etal2002}, which widely agree with each other).

Figure \ref{sfh-hz} shows the star formation history, derived by \citet{harris+zaritsky2009}.
The star formation history of the fields within the regions of \citet{massey2002} (open circles in Fig. \ref{selection}) in which star clusters have been observed is shown as the solid line, with the grey area being its uncertainty.
This partial star formation history follows the total star formation history (for the whole area of the Large Magellanic Cloud, dotted line) at about half the star formation rate.

\section{Comparison of the Methods}\label{sec_comparison}

\begin{figure}
\includegraphics[width=8cm]{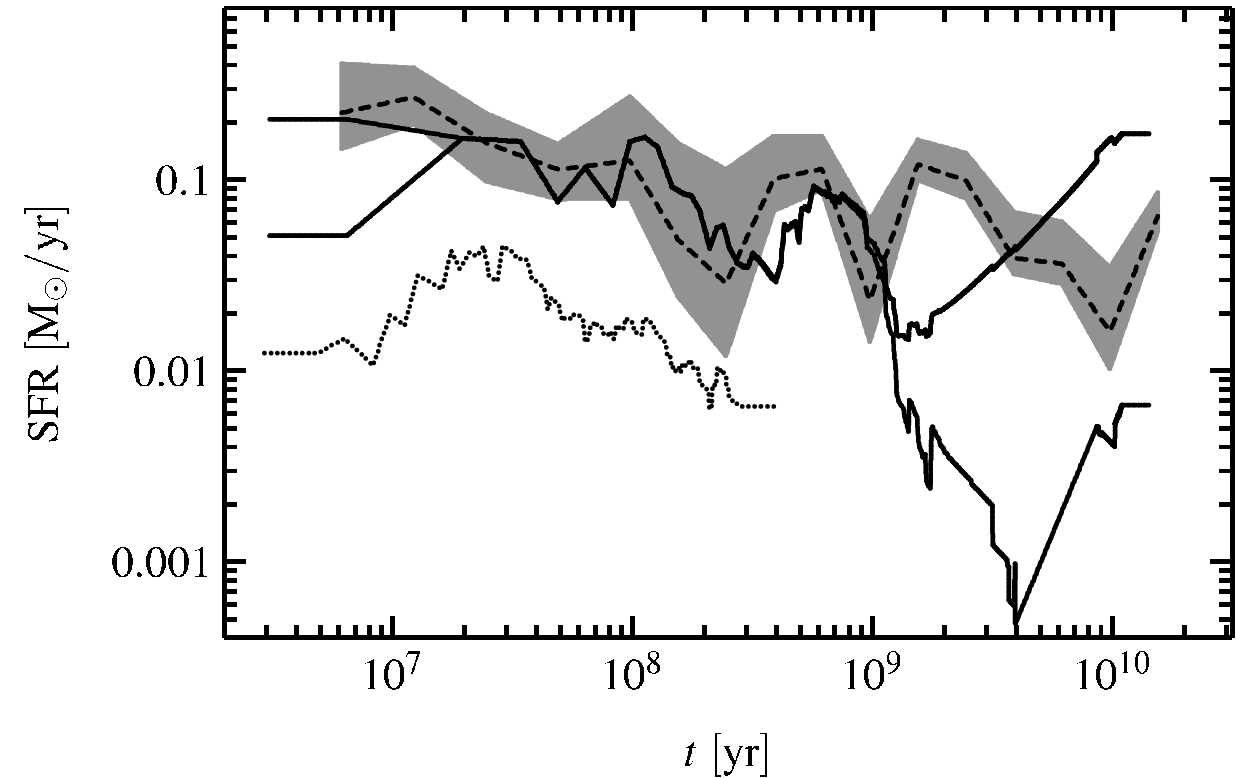}
\caption{\label{sfh-comparison}
Comparison of the results for the star formation history in the Large Magellanic Cloud.
The dashed line within the grey shaded region (its uncertainty) is derived from colour magnitude diagrams for the parts lying in the \citet{massey2002} fields (as in Fig. \ref{sfh-hz}, Sec. \ref{sec_cmd}).
The thick solid lines are the solution using the most massive clusters, with the upper and lower limit for older ages, assuming no cluster formation or the maximum non detectable cluster mass in empty age bins (Fig. \ref{sfhmmax} bottom panel, Sec. \ref{sec_mostmassive}).
The lowest dotted line is derived from the total star cluster mass (time variable $M_c$(middle choice from top panel of Fig. \ref{sfhdirect}, $M_l=100\ \Msun$, $\beta=2$, for details see Sec. \ref{sec_total}).
Cluster masses have been dynamically back-evolved using $t_4$ 8 Gyr.
}
\end{figure}

In Figure \ref{sfh-comparison} we summarise the solutions for the star formation history of the Large Magellanic Cloud derived from colour-magnitude diagrams, the most massive star cluster or the total star cluster mass.
As before in Fig. \ref{sfh-hz} the dashed line within the grey shaded area gives the star formation history with uncertainty derived from colour-magnitude diagrams within the \citet{massey2002} fields.
The thick solid lines give the upper and lower limits for the star formation history derived from the most massive star clusters (Sec. \ref{sec_mostmassive}) and the dotted line gives the amount of star formation in star clusters (Sec. \ref{sec_total}).

The comparison of the $M_\mathrm{max}$ estimate for the star formation history with the result from the CMD shows good agreement for ages up to $\approx 1$ Gyr, choosing the $M_\mathrm{max}$ solution which includes R136.
The absolute value for the star formation rate, derived from the most massive clusters, is at the level of the CMD star formation rate within the \citet{massey2002} fields.
However, as the most massive cluster-method is intended to give the star formation rate {\it of an entire galaxy}, and the spatial coverage of the star clusters contains most of the area recently active in star formation, it is perhaps more appropriate to compare to the galaxy-wide star formation rate.
In this case the star formation rate would be underestimated by a factor of 2 (compared to the dotted line in Fig. \ref{sfh-hz}) and the normalisation in eq. \ref{sfrdeltat} would need adjusting.
As long as the spatial coverage is incomplete it is unfortunately impossible to disentangle inappropriate normalisation and effects of spatial incompleteness.

Compared to the CMD solution the most massive cluster solution shows an offset of the peaks in star formation.
This is perhaps accounted for by the moving averaging window, although different isochrone sets could also account for this.
If the ages of the star clusters could be more accurately determined the moving window could be reduced, by at the same time including not only the most massive star cluster but also the second, third etc. most massive and thus keeping the sample of data points used large enough for the statistically necessary averaging.
An investigation in this direction is beyond the scope of this work.

We turn now to a the comparison of the star formation rate derived from the total mass with the CMD and $M_\mathrm{tot}$ solution.
For the very youngest ages ($< 10$--$20$ Myr) the $M_\mathrm{tot}$ star formation history shows an increase of the star formation rate with age, in contrast to the {\it decrease} of the CMD solution.
This is because very young clusters are in a class of objects not selected by \citet{hunter-etal2003} and thus the used data set is incomplete for very young ages.
Within the age range where we can assume that the $M_\mathrm{tot}$ solution is based on a complete data set (both in object selection and lower mass completeness), from $\approx 10$--$20$ Myr to $\approx$ 200 Myr, the $M_\mathrm{tot}$ solution shows the same structure as the CMD and $M_\mathrm{max}$ solution, but at a lower star formation rate.
The fraction of star formation in star clusters, i.e. the ratio between the CMD curve and the $M_\mathrm{tot}$ curve, appears for the whole age range to be at a 10--20\% level.
This is caused by either the formation of only a fraction of stars in star clusters, or, when assuming that all stars form in custers, by the dissolution of star clusters caused by the presumably violent transition from their gas-embedded to the gas-free state.
The second explanation could in principle be detected in the comparison, it should lead to a $M_\mathrm{tot}$ star formation rate that is identical to the CMD star formation rate at very young ages (the embedded cluster phase).
However, as the data set is not complete at the youngest ages a distinction between the alternative explanations is not possible.

For ages older than 1 Gyr the number of detected star clusters is very small, and their distribution shows many gaps.
Thus no large amount of information about the star formation history can be extracted from the star clusters.
The $M_\mathrm{max}$ method gives an upper and lower limit for the star formation history, either assuming no star cluster formation in the gaps (lower bold line in Fig. \ref{sfh-comparison}) or assuming the mass of the detection limit as the upper limit for the mass of a cluster that could have formed in an gap (upper line).
As the mass associated with the detection limit increases with increasing age, the upper limit for the star formation history shows also an increase of the star formation rate with age.
The uncertainties introduced in the $M_\mathrm{max}$ solution by stochastical scatter and mass errors are small compared to the effect of missing data.
As there are some clusters with very high initial masses and ages around 10 Gyr one would conclude that there was a very high star formation activity, which is visible in the lower limit provided by the $M_\mathrm{max}$ solution.

Given the good agreement of the $M_\mathrm{max}$ and CMD method at younger ages one would expect that the CMD solution falls in the region between the two $M_\mathrm{max}$ limits.
However, the CMD solution shows at an age of 2 Gyr a peak in the star formation rate which is nearly a factor of 10 higher than the upper limit for the star formation rate following form $M_\mathrm{max}$.
Although there are some clusters at 2 Gyr with $\approx 10^5$ \Msun (which lead to some kind of ``peak'' there in the lower limit of the $M_\mathrm{max}$ method), many more would be needed to produce a signal.
It also seems to be odd that there are essentially no clusters with masses between $10^4\ \Msun$ (the observational completeness limit, cf. Fig. \ref{sfhmmax}) and $10^5\ \Msun$ (the mass of the observed clusters), a mass range which should be populated assuming a normal cluster mass function.

If star formation is coupled to star cluster formation, as assumed for the $M_\mathrm{max}$ method, and as shown by the agreement for ages $<$ 1 Gyr, there should actually be also a large number of star clusters with ages of $\approx 2 $ Gyr.
There are perhaps several explanations for these missing clusters: they are located outside the observed fields, or a very different mode of star formation was active, where no massive clusters are produced, or a cluster destruction mechanism which only affected the missing clusters \citep[e.g. interactions between the Large and Small Magellanic Cloud as suggested by ][]{bekki-etal2004}.

\section{Summary and Conclusions}\label{sec_summary}

We compared the results for the star formation history of the Large Magellanic Cloud derived from its star cluster population, either following the method of \citet{maschberger+kroupa2007}, using the most massive clusters only, or using the total mass of the whole mass range of clusters (but then only for the most recent 400 Myr),
with the results derived from colour-magnitude diagrams by \citet{harris+zaritsky2009}.

We found that the results using the most massive clusters, both the absolute value for the star formation rate and the structure of the star formation history, agree well for the first Gyr with the star formation history derived from a CMD.
For ages larger than 1 Gyr only a small number of clusters is detected, so that only a lower and upper limit of the star formation history can be given with the $M_\mathrm{max}$ method.
The CMD solution does not fall within these limits  between 1 and 3 Gyr, but shows a peak with a higher star formation rate.
This implies that the number of detected star clusters is too small compared to the expectations from the star formation rate following from CMDs.
One possibility to resolve this discrepancy would be that additional clusters are contained in the area which is not observationally covered.

Furthermore we derived the star formation history using all available star clusters, which, however, is only feasible for the most recent 20--400 Myr.
 The shape if this star formation history agrees with the CMD and $M_\mathrm{max}$ results, albeit with absolute values for the star formation rate a factor $\approx$ 10 lower.
This implies that the fraction of star formation in (presumably bound, open) star clusters after gas expulsion is at a 10--20\% level.
Alternatively, this means (assuming that all stars form in a clustered way), that star clusters have an infant mortality of 80--90\%.

Our results show that star clusters are a powerful means to investigate the star formation history of a galaxy and invite further investigation in that direction.

\section{Acknowledgements}
We wish to thank Richard de Grijs and Peter Anders for providing us with their data of the star clusters and Cathie Clarke for valuable comments on the manuscript.
Th. M. acknowledges funding in Cambridge through {\sc constellation}, an European Commission FP6 Marie Curie Research Training Network, and the Stellar Populations and Dynamics Research Group at the Argelander-Institut f{\"ur} Astronomie at Bonn University.

\bibliographystyle{mn2e}
\bibliography{cluster}

\bsp

\label{lastpage}

\end{document}